\documentclass[aip,jcp,preprint,numerical,amsmath,amssymb,amsfonts]{revtex4-1}
\usepackage{bm,graphicx,xcolor,hyperref,rotating,lineno,mathtools,multirow,makecell,placeins,physics}
\usepackage[version=4]{mhchem}
\usepackage{hyperref}
\hypersetup{
    colorlinks=true,
    linkcolor=blue,
    filecolor=blue,
    urlcolor=blue,
    citecolor=blue
}

\usepackage[normalem]{ulem}

\newcommand{\josh}[1]{\textcolor{black}{#1}}

\newcommand{\DNO}{\Delta\text{NO}}

\newcommand{\br}{\mathbf{r}}
\newcommand{\bx}{\mathbf{x}}
\newcommand{\A}{\text{A}}
\newcommand{\ON}[1]{n_{#1}}
\newcommand{\VN}[1]{h_{#1}}
\newcommand{\NO}[1]{\phi_{#1}}
\newcommand{\NSO}[1]{\chi_{#1}}
\newcommand{\RRDM}{\Tilde{\Gamma}}
\newcommand{\rRRDM}{\Gamma}
\newcommand{\RDM}{\Tilde{\gamma}}

\newcommand{\up}{\uparrow}
\newcommand{\dw}{\downarrow}
\newcommand{\dyn}{\text{dyn}}
\newcommand{\stat}{\text{stat}}
\newcommand{\cum}{\text{cum}}
\newcommand{\pair}{\text{pair}}
\newcommand{\HSC}{\text{HSC}}
\newcommand{\hc}{H}

\newcommand{\Occ}{\mathcal{O}}
\newcommand{\Virt}{\mathcal{V}}
\newcommand{\AO}{\mathcal{A}_o}
\newcommand{\AV}{\mathcal{A}_v}
\newcommand{\Act}{\mathcal{A}}

\newcommand{\res}[2]{r_{#1}^{#2}}
\newcommand{\damp}[2]{d_{#1}^{#2}}
\newcommand{\damps}[2]{s_{#1}^{#2}}
\newcommand{\amp}[2]{t_{#1}^{#2}}

\makeatletter
\newcommand\footnoteref[1]{\protected@xdef\@thefnmark{\ref{#1}}\@footnotemark}
\makeatother

\newcommand{\LCPQ}{Laboratoire de Chimie et Physique Quantiques (UMR 5626), Universit\'e de Toulouse, CNRS, UPS, France}
\newcommand{\UWig}{Department of Chemistry, University of Winnipeg, Winnipeg, Manitoba, R3B 2G3, Canada}
\newcommand{\UMan}{Department of Chemistry, University of Manitoba, Winnipeg, Manitoba, R3T 2N2, Canada}

\begin{document}

\title{Capturing static and dynamic correlation with $\DNO$-MP2 and $\DNO$-CCSD}
\author{Joshua W. Hollett}
\email[Corresponding author: ]{j.hollett@uwinnipeg.ca}
\affiliation{\UWig}
\affiliation{\UMan}
\affiliation{\LCPQ}
\author{Pierre-Fran\c{c}ois Loos}
\affiliation{\LCPQ}

\begin{abstract}
The $\DNO$ method for static correlation is combined with second-order M{\o}ller-Plesset perturbation theory (MP2) and coupled-cluster singles and doubles (CCSD) to account for dynamic correlation.  The MP2 and CCSD expressions are adapted from finite-temperature CCSD, which includes orbital occupancies and vacancies, and expanded orbital summations.  Correlation is partitioned with the aid of damping factors incorporated into the MP2 and CCSD residual equations.  Potential energy curves for a selection of diatomics are in good agreement with extrapolated full configuration interaction results (exFCI), and on par with conventional multireference approaches.  
\end{abstract}

\maketitle

\section{Introduction}
\label{sec:intro}
The correlation problem persists. To state it simply; how does one adequately account for electron correlation with a minimal amount of effort?  Its persistence is ensured by the latter condition. This continual search for an efficient treatment of electron correlation is driven by the need to treat larger and more complex systems with increased accuracy.  A common strategy of potential solutions is the partitioning of the problem into different types of correlation; static and dynamic,\cite{Mok1996,Handy2001,Cremer2001,Becke2013,Crittenden2013,Tsuchimochi2014,Wallace2014,NOF1,RamosCordoba2016,BenavidesRiveros2017,ViaNadal2019} long-range and short-range,\cite{Fromager2007,Toulouse2009,Janesko20091,Chai2009,Stoyanova2013} \textit{etc.}  Partitioning the correlation problem into static and dynamic correlation, or strong and weak correlation, or multireference and ``the rest'', is a popular and effective strategy that generally provides a qualitative, and sometimes quantitative, model for particularly challenging electronic structure problems. The price of the success of such models is relatively expensive calculations, often combined with the non-trivial definition of active spaces that requires both chemical intuition and trial-and-error. Through the reformulation of these problems in terms of alternative models of electronic structure, a deeper and more ``physical'' understanding of correlation partitioning can be achieved while providing another tool for the study of complex multireference systems.

A two-tiered wave function based approach to static and dynamic correlation is a relatively old idea.\cite{Cizek1969,Jeziorski1981,Andersson1990,Pulay2011,Evangelista2018} The general strategy of manually (or automatically\cite{Sayfutyarova2017,Bao2018}) defining an active space, optimizing a multireference wave function, and then applying some form of post-Hartree-Fock electron correlation method, is the basis of a multitude of multireference electronic structure models.\cite{Szalay2012,Lyakh2012,Kohn2013,Evangelista2018}  These models have continually evolved over the decades, and prominently include CASPT2,\cite{Andersson1990,Andersson1992,Shiozaki2011} NEVPT,\cite{Angeli2001a,Angeli2001b,Angeli2002} MRCC, \cite{Cizek1969,Jeziorski1981,Hoffmann1999,Lyakh2005} and NOCI. \cite{Thom2009,Yost2013,Sundstrom2014,Burton2019} These methods are essentially the default for studying systems with low-lying excited states (\textit{e.g.},~conical intersections), largely because conventional density functional methods often fail to properly model such systems.  An emerging alternative to these approaches, particularly for describing the multireference aspect, is cumulant functional methods [\textit{e.g.},~density-matrix functional theory (DMFT)\cite{Gilbert1975,Zumbach1985,Kutzelnigg2006,Sokolov2013,Mentel2014,vanMeer2018,Schilling2018,Schmidt2019} and natural orbital functional theory (NOFT)\cite{Muller1984,Goedecker1998,Rohr2008,PNOF0a,PNOF0b,PNOF1to5,PNOF6,NOFMP2a})].  Recently, a two-tiered approach as seen in wave function approaches was devised for NOFT by Piris, NOF-MP2.\cite{NOFMP2a,NOFMP2b} Other than NOF-MP2, the combination of a cumulant functional for static correlation and post-Hartree-Fock theories for dynamic correlation is unexplored.

Upon its inception, the $\DNO$ method\cite{NOF1} involved employing a cumulant functional to account for static correlation (or multireference character) in conjunction with an on-top density functional for dynamic correlation.  The on-top density functional is applied directly to the statically correlated $\DNO$ two-electron density matrix (2-RDM), therefore the method for treating dynamic correlation can be easily substituted. \josh{Recently, multiple formulations of finite-temperature coupled-cluster approximations have been introduced,\cite{Margraf2018,Hummel2018} including coupled-cluster singles and doubles (FT-CCSD)\cite{White2018} by White and Chan, which is similar to thermal cluster cumulant theory.\cite{Sanyal1992,Sanyal1993,Mandal2003} In such an approach, orbitals are thermally populated according to a Fermi-Dirac distribution, therefore there are non-integer electron occupancies and vacancies (holes).} Similar formulations also exist for second-order M{\o}ller-Plesset perturbation theory (FT-MP2).\cite{Kobayashi2015,Santra2017,Mandal2003}  The finite-temperature formulations of post-Hartree-Fock approaches present an ideal framework for treating the dynamic correlation of a multireference (or statically correlated) 2-RDM obtained from $\DNO$, or elsewhere.

The method presented here involves combining $\DNO$ for static correlation with MP2 or CCSD for dynamic correlation, by exploiting aspects of the finite-temperature formulation.  The combination is made possible by introducing a $\Delta$-dependent damping factor in the leading term of the MP2 or CCSD residuals, which modifies the occupancy-occupancy, vacancy-vacancy, and occupancy-vacancy pairs according to the amount of static correlation present. The $\DNO$ method is introduced in Subsec.~\ref{subsec:deltano}, the modified MP2 and CCSD equations are described in Subsecs.~\ref{subsec:mp2} and \ref{subsec:ccsd}, and the damping factors are defined in Appendix \ref{app:damp}.  The implementation of the method is described in Sec.~\ref{sec:method} and results for the dissociation of some diatomics are presented and discussed in Sec.~\ref{sec:results}.  Finally, some conclusions regarding the current implementation and some future directions are discussed in Sec.~\ref{sec:conc}. Atomic units are used throughout unless stated otherwise.

\section{Theory}
\label{sec:theory}

\subsection{$\DNO$ method}
\label{subsec:deltano}

Cumulant functional theory (CFT) is based on the cumulant expansion of the exact two-electron reduced density matrix (2-RDM) in terms of the one-electron reduced density matrix (1-RDM) and occasionally other variables.\cite{Kutzelnigg1999}  The 2-RDM can be defined in terms of the $N$-electron wave function,
\begin{multline} 
	\label{eq:2RDM}
	\RRDM(\bx_1,\bx_2,\bx_1',\bx_2') = \frac{N(N-1)}{2} \int \Psi^*(\bx_1',\bx_2',\bx_3,\dots,\bx_N) 
	\\ 
	\times \Psi(\bx_1,\bx_2,\bx_3,\dots,\bx_N) d\bx_3 \dots d\bx_N,
\end{multline}
where $\bx = (\br,\omega)$ represents both the spatial and spin coordinates of an electron. The 1-RDM follows from the 2-RDM via integration of the coordinates of one of the electrons,
\begin{equation} \label{eq:1RDM}
	\RDM(\bx,\bx') = \frac{2}{N-1} \int \tilde{\Gamma}(\bx,\bx_2,\bx',\bx_2) d\bx_2.
\end{equation}
The cumulant expansion of the 2-RDM can be written as
\begin{equation}
	\RRDM(\bx_1,\bx_2,\bx_1',\bx_2') 
	= \RRDM^{(0)}(\bx_1,\bx_2,\bx_1',\bx_2') + \RRDM_\cum(\bx_1,\bx_2,\bx_1',\bx_2'),
\end{equation}
where the zeroth-order term of the expansion, $\RRDM^{(0)}$, is expressed solely in terms of the 1-RDM,
\begin{equation}
	\RRDM^{(0)}(\bx_1,\bx_2,\bx_1',\bx_2') 
	= \frac{1}{2} \qty[ \RDM(\bx_1,\bx_1') \RDM(\bx_2,\bx_2') - \RDM(\bx_1,\bx_2') \RDM(\bx_2,\bx_1') ].
\end{equation}
The general form of the cumulant, $\RRDM_\text{cum}$, for an $N$-electron system is unknown, and present CFT methods are distinguished by how they approximate this term.  When $\RRDM_\text{cum}$ is constructed exclusively from the natural orbitals (NOs), $\qty{ \NO{p} }$, and their occupancies, $\qty{ \ON{p} }$, (which are the eigenfunctions and eigenvalues of the 1-RDM, respectively) a natural orbital functional (NOF) is the result.  For notational convenience, we also define natural vacancies as $\VN{p} = 1 - \ON{p}$ and assume real-valued NOs.  Unlike NOFs, the $\DNO$ method uses electron transfer variables, $\qty{ \Delta_{me} }$, which correspond to the amount of electron occupancy transferred from an ``occupied'' active orbital $\NO{m}$ to a ``virtual'' active orbital $\NO{e}$. Note that ``occupied'' and ``virtual" designations refer to the ground-state Hartree-Fock electron configuration \cite{NOF1} (see Table \ref{tab:index} for orbital labelling). 

\begin{table}
	\caption{\label{tab:index} Orbital index key for $\Delta$NO, MP2 and CCSD.}
	\begin{ruledtabular}
	\begin{tabular}{lllc}
			indices & orbitals & trait & set label \\
		\hline
		$p,q,r,s$ 	&	all				&			& 		\\
		$i,j,k,l$	&	occupied		&	$\ON{i} \ne 0$	& $\Occ$	\\
		$a,b,c,d$	&	virtual			&	$\VN{a} \ne 0$	& $\Virt$ 	\\
		$m,n$		&	active occupied	&	$\ON{m} < 1$	& $\AO$		\\
		$e,f$		&	active virtual	&	$\VN{e} < 1$	& $\AV$		
	\end{tabular}
	\end{ruledtabular}
\end{table}
In $\DNO$, the occupancies are defined in terms of these variational $\qty{ \Delta_{me} }$,
\begin{align}
\label{eq:delta}
	\ON{m} & = 1 - \sum_e \Delta_{me},
	&
	\ON{e} & = \sum_m \Delta_{me}.
\end{align}
Further distinguishing the $\DNO$ functional from NOFs, or other cumulant functionals, is that the transfer of electrons occurs between a relatively small set of active occupied, $ \AO = \qty{ \NO{m} }$, and virtual, $ \AV = \qty{ \NO{e} }$, orbitals.  This is because the $\DNO$ cumulant functional is designed to capture only static correlation.

For this work, it is useful to describe the spinless, spin-resolved, $\DNO$ 2-RDM.  In general, the spinless 2-RDM is obtained by integrating over the spin of the two electrons,
\begin{equation} \label{eq:2rdmspinless}
	\rRRDM(\br_1,\br_2,\br_1',\br_2') 
	= \iint \left. \RRDM(\bx_1,\bx_2,\bx_1',\bx_2') \right|_{\substack{\omega_1' = \omega_1\\ \omega_2' = \omega_2}} d\omega_1 d\omega_2.
\end{equation}
The result can then be resolved into the components associated with different spin-pairs,
\begin{equation} 
\label{eq:2rdmspincomp}
\begin{split} 
	\rRRDM(\br_1,\br_2,\br_1',\br_2') 
	& =
	\rRRDM^{\up\up}(\br_1,\br_2,\br_1',\br_2') + \rRRDM^{\dw\dw}(\br_1,\br_2,\br_1',\br_2') 
	\\
	& + \rRRDM^{\up\dw}(\br_1,\br_2,\br_1',\br_2') + \rRRDM^{\dw\up}(\br_1,\br_2,\br_1',\br_2').
\end{split}
\end{equation}
Furthermore, the 2-RDM can also be expanded in the basis of the NOs,
\begin{equation}
	\rRRDM(\br_1,\br_2,\br_1',\br_2') = \sum_{pqrs} \rRRDM_{pqrs} \NO{p}(\br_1')\NO{q}(\br_2')\NO{r}(\br_1)\NO{s}(\br_2).
\end{equation}
For a closed-shell system, the zeroth-order term of the cumulant expansion becomes
\begin{subequations}	\label{eq:Gamma01RDM}
\begin{align}
	\qty( \rRRDM^{(0),{\sigma\sigma}})_{pqrs} 
	& = \frac{\ON{p} \ON{q}}{2} \delta_{pr}^{qs},
	\\
	\qty( \rRRDM^{(0),\sigma\sigma'})_{pqrs}  
	& = \frac{\ON{p} \ON{q}}{2} \delta_{pr} \delta_{qs},
\end{align}
\end{subequations}
where $\delta_{pr}^{qs} = \delta_{pr} \delta_{qs} - \delta_{ps} \delta_{qr}$, $\ON{p} = \ON{p}^\up = \ON{p}^\dw$ and $\sigma$, $\sigma'$ = $\up$ or $\dw$.

The $\DNO$ cumulant consists of three terms,
\begin{equation}
	\rRRDM^{\DNO}_\cum = \rRRDM^{\DNO}_\pair + \rRRDM^{\DNO}_\stat + \rRRDM^{\DNO}_\HSC,
\end{equation}
a pair correction term, $\Gamma^{\DNO}_\pair$, a static correlation term, $\Gamma^{\DNO}_\stat$, and a high-spin correction term, $\Gamma^{\DNO}_\HSC$, where each can be decomposed into its spin-pair components. 

For non-integer occupancies, $\rRRDM^{(0)}$ [see Eq.~\eqref{eq:Gamma01RDM}] does not integrate to the total number of electron pairs, $N(N-1)/2$.  The pair correction term, $\rRRDM^{\DNO}_\pair$, ensures the total 2-RDM integrates to this number for any $\qty{ \Delta_{me} }$, and is given as
\begin{subequations}	
\begin{align}
\label{eq:Gammapair}
	\qty( \rRRDM_\pair^{\DNO,\sigma\sigma} )_{pqrs} 
	& = \frac{\Delta_{pq}(\ON{q} - \ON{p} -\Delta_{pq}) - \eta_{pq}}{2} \delta_{pr}^{qs},
\\
\begin{split}
	\qty( \rRRDM_\pair^{\DNO,\sigma\sigma'} )_{pqrs} 
	& = \frac{\ON{p} \VN{p}}{2} \delta_{pq} \delta_{pr} \delta_{qs}
	\\
	& + \frac{\Delta_{pq}(\ON{q} - \ON{p} -\Delta_{pq}) - \eta_{pq}}{2} \delta_{pr} \delta_{qs},
\end{split}
\end{align}
\end{subequations}
where
\begin{equation}	
\label{eq:eta}
		\eta_{pq} = 	
		\begin{cases}
			\sum_r \Delta_{pr} \Delta_{qr}, & \text{if }p\ne q \land (\NO{p}, \NO{q}) \in \AO \\
							& \text{or }p\ne q \land (\NO{p}, \NO{q}) \in \AV,	\\
			0, & \text{otherwise},
		\end{cases}
\end{equation}
and $\Delta_{pq} = -\Delta_{qp}$.

In the framework of $\DNO$, static correlation is captured by transferring opposite-spin electron pairs from the same active occupied NO, $\NO{m}$, to the same active virtual NO, $\NO{e}$.  This recovers the same intrapair correlation as the $2n$-tuple excitations of a seniority-zero configuration interaction wave function,\cite{Bytautas2011,Polemans2015} for which excitations are performed only within a relatively small active space.  

The static correlation term of the cumulant is written as
\begin{subequations}
\begin{align}
	\qty( \rRRDM_\stat^{\DNO,\sigma\sigma} )_{pqrs} & = 0,
	\\
	\qty( \rRRDM_\stat^{\DNO,\sigma\sigma'} )_{pqrs} & = \frac{\zeta_{pr} - \tau_{pr}}{2} \delta_{pq} \delta_{rs},
\end{align}
\end{subequations}
where
\begin{equation}	
\label{eq:zeta}
	\zeta_{pq} = 	
		\begin{cases}
			\sum_r \sqrt{\Delta_{pr} \Delta_{qr}}, & \text{if }p\ne q \land (\NO{p}, \NO{q}) \in \AO \\
							& \text{or }p\ne q \land (\NO{p}, \NO{q}) \in \AV,
			\\
			0, & \text{otherwise},
		\end{cases}
\end{equation}
and
\begin{equation}	
\label{eq:tau}
	\tau_{pq} = 	
		\begin{cases}
			\sqrt{\ON{p} \Delta_{pq}}, & \text{if }\NO{p} \in \AO \land \NO{q} \in \AV,
			\\
			\sqrt{\ON{q} \Delta_{qp}}, & \text{if }\NO{p} \in \AV \land \NO{q} \in \AO,
			\\
			0, & \text{otherwise}.
		\end{cases}
\end{equation}

\begin{figure}
	\begin{center}
	\includegraphics[width=0.7\textwidth]{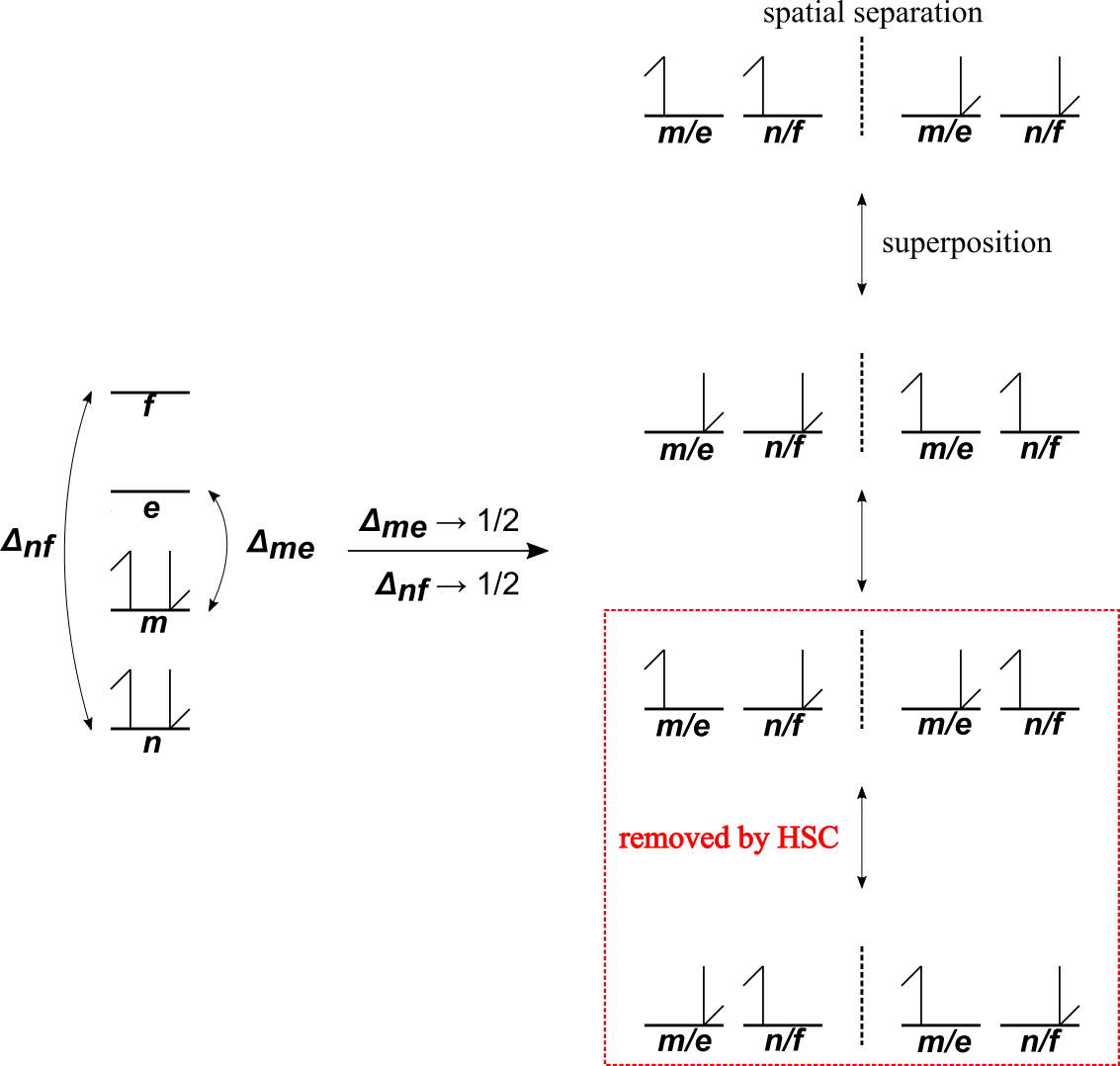}
	\caption{\label{fig:HSC} Diagrammatic representation of the $\DNO$ high-spin correction (HSC) for two statically correlated electron pairs. Without the HSC, at complete static correlation ($\Delta = 1/2$), the $\DNO$ 2-RDM would contain a superposition of four different spin configurations about a spatial separation [dashed line] (\textit{e.g.},~dissociated atoms).  The HSC removes the ``low-spin" configurations, leaving only the ``high-spin" configurations.}
	\end{center}
\end{figure}

Like a seniority-zero wave function, no parallel-spin correlation is included in the static correlation term. However, the high-spin correction (HSC) includes interpair, opposite- and parallel-spin, correlation that is not present in a seniority-zero wave function.  This correlation is necessary for the proper dissociation of multiple bonds into high-spin fragments, and for the static correlation of multiple electron pairs in general.\cite{vanMeer2018}  The HSC term is written as
\begin{subequations}	
\label{eq:GammaHSC}
\begin{align}
	\qty( \rRRDM_\HSC^{\DNO,\sigma\sigma} )_{pqrs} & = \frac{\kappa_{pq}}{2} \delta_{pr}^{qs},
	\\
	\qty( \rRRDM_\HSC^{\DNO,\sigma\sigma'} )_{pqrs} & = -\frac{\kappa_{pq}}{2} \delta_{pr}\delta_{qs},
\end{align}
\end{subequations}
where
\begin{equation}	
\label{eq:kappa}
	\kappa_{pq} = 	
		\begin{cases}
			\sum_{\substack{r \ne s \\ (r \ne q) \\ (s \ne p)}} \tau_{pr} \tau_{qs}, & \text{if } p\ne q \land (\NO{p}, \NO{q}) \in \Act,
			\\
			0, & \text{otherwise}.
		\end{cases}
\end{equation}
The HSC reduces the pair density between opposite-spin electrons, while increasing the pair density between parallel-spin electrons, of separate statically correlated electron pairs, as their static correlation increases.  This is illustrated in Fig.~\ref{fig:HSC} for two statically correlated electron pairs.
Taking \ce{N2} dissociation as an example, the HSC ensures that the spin-up electrons of each of the three statically correlated pairs (triple bond) simultaneously appear on one atom while the spin-down electrons appear on the other, resulting in a superposition of the two high-spin fragment possibilities.  Without the correction, the electrons of each statically correlated pair would encounter an average of both parallel and opposite-spin electrons from the other statically correlated pairs (\textit{i.e.},~spin-averaged).

The total $\DNO$ energy follows simply from the 2-RDM,
\begin{equation}	
\label{eq:edeltano}
	E^{\DNO} = E^{(0)} + E^{\DNO}_\cum,
\end{equation}
where the zeroth-order 1-RDM energy also includes the one-electron, kinetic and electron-nucleus attraction, energy in addition to the two-electron energy associated with the zeroth-order term of the cumulant expansion, $\Gamma^{(0)}$. For a closed-shell system, the zeroth-order 1-RDM energy, in terms of NOs and occupancies, is given as
\begin{equation}	
\label{eq:e01rdm}
	E^{(0)} = 2 \sum_p \ON{p} \hc_{p} + \sum_{pq} \ON{p} \ON{q} \qty( 2J_{pq} - K_{pq} ),
\end{equation}
where
\begin{subequations}	
\label{eq:hJK}
\begin{align}
	\hc_{p} & = \int \NO{p}(\br) \qty(-\frac{\nabla^2}{2} - \sum_\A \frac{Z_\A}{r_\A} ) \phi_p(\br) d\br,
	\\
	J_{pq} & = \iint \frac{\NO{p}(\br_1)\NO{q}(\br_2)\NO{p}(\br_1)\phi_q(\br_2)}{r_{12}} d\br_1 d\br_2,
	\\
	K_{pq} & = \iint \frac{\NO{p}(\br_1)\NO{q}(\br_2)\NO{q}(\br_1)\NO{p}(\br_2)}{r_{12}} d\br_1 d\br_2,
\end{align}
\end{subequations}
are the usual one-electron (kinetic and nuclear attraction) and two-electron (Coulomb and exchange) integrals over NOs.
The cumulant energy is given as
\begin{equation}	
\label{eq:eq}
	E^{\DNO}_\text{cum} = E^{\DNO}_\pair + E^{\DNO}_\stat + E^{\DNO}_\HSC + E^{\DNO}_\dyn,
\end{equation}
with components defined as follows
\begin{subequations}	
\begin{align}	
\label{eq:Epair}
\begin{split}	
	E^{\DNO}_\pair 
	& = \sum_p \ON{p} \VN{p} J_{pp}
	\\
	& + \sum_{pq} \Delta_{pq} \qty(\ON{q} - \ON{p} - \Delta_{pq} )\qty( 2 J_{pq} - K_{pq} )
	\\
	& - \sum_{pq} \eta_{pq} \qty( 2J_{pq} - K_{pq} ),
\end{split}
\\
\label{eq:estat}
	E^{\DNO}_\stat 
	& = \sum_{pq}\left( \zeta_{pq} - \tau_{pq} \right) L_{pq},
	\\
	\label{eq:hsc}
	E^{\DNO}_\HSC 
	& = -\sum_{pq} \kappa_{pq} K_{pq},
\end{align}
\end{subequations}
where the time-inversion exchange energy integrals are
\begin{equation}
	L_{pq} = \iint \frac{\NO{p}(\br_1)\NO{p}(\br_2)\NO{q}(\br_1)\NO{q}(\br_2)}{r_{12}} d\br_1 d\br_2.
\end{equation}
The HSC energy appears simplified in comparison to the 2-RDM term [see Eq.~\eqref{eq:GammaHSC}]. This is because the Coulomb repulsion terms cancel due to the equivalence of the spin-up and spin-down NOs.  The dynamic correlation energy, $E^{\DNO}_\dyn$, was defined previously in terms of an on-top density functional.\cite{NOF1} In the present study the dynamic correlation energy is provided via MP2 or CCSD, \textit{i.e.},
\begin{equation}
	E^{\DNO}_\dyn = E^{\DNO}_\text{MP2/CCSD}.
\end{equation}

\subsection{CCSD for $\DNO$}
\label{subsec:ccsd}

Recently, White and Chan introduced a finite-temperature formulation of the coupled-cluster singles and doubles method (FT-CCSD).\cite{White2018}  The method is formulated in terms of imaginary time, which is integrated from $0$ to $\beta$, where $\beta$ is the inverse temperature. The authors state that at zero temperature, the FT-CCSD amplitudes, and consequently the energy, converge to the usual non-temperature dependent CCSD values.  In that case, the electron occupancies, which are determined by a Fermi-Dirac distribution, would collapse to their normal Aufbau (Hartree-Fock ground state) values. For $\DNO$, the occupancies are not those of Aufbau or the Fermi-Dirac distribution, nevertheless it is assumed here that aspects of the FT-CCSD formulation are still valid.  In their article, White and Chan outline how to convert CC equations (\textit{i.e.},~residuals) to FT-CC equations.  The equations presented here are formulated by taking the CCSD equations of Stanton \textit{et al.}\cite{Stanton1991}~and applying the instructions from White and Chan to include occupancies and vacancies (holes).  The necessary instructions (paraphrased) being: (i) for each contraction, sum over all orbitals instead of just occupied or virtual orbitals, and (ii) include an occupancy or vacancy with each index not associated with an amplitude.  
Application of these instructions to the residual ($\res{i}{a}$ and $\res{ij}{ab}$) equations of Stanton \textit{et al.}~gives
\begin{equation} 
\label{eq:r1ccsd}
\begin{split} 
	\res{i}{a}(\text{CCSD}) 
	& = \ON{i} \VN{a} \damps{i}{a} F_{ia} + \sum_c \amp{i}{c} \mathcal{F}_{ac} - \sum_k \amp{k}{a} \mathcal{F}_{ki}
	\\
	& + \sum_{kc} \amp{ik}{ac} \mathcal{F}_{kc} - \sum_{kc} \amp{k}{c} \mel{ka}{}{ic} \ON{i} \VN{a} 
	\\
	& - \frac{1}{2} \sum_{kcd} \amp{ik}{cd} \mel{ka}{}{cd} \VN{a} 
	\\
	& - \frac{1}{2} \sum_{klc} \amp{kl}{ac} \mel{lk}{}{ci} \ON{i},
\end{split}
\end{equation}
and
\begin{equation} 
\label{eq:r2ccsd}
\begin{split} 
	\res{ij}{ab}(\text{CCSD}) 
	& = \ON{i} \ON{j} \VN{a} \VN{b} \damp{ij}{ab} \mel{ij}{}{ab}
	\\
	& + P_{ab} \sum_c \amp{ij}{ac} \qty( \mathcal{F}_{bc} - \frac{1}{2} \sum_k \amp{k}{b} \mathcal{F}_{kc} )
	\\
	& - P_{ij} \sum_k \amp{ik}{ab} \qty( \mathcal{F}_{kj} + \frac{1}{2} \sum_c \amp{j}{c} \mathcal{F}_{kc} )
	\\
	& + \frac{1}{2} \sum_{kl} \tau_{kl}^{ab} \mathcal{W}_{klij} +\frac{1}{2} \sum_{cd} \tau_{ij}^{cd} \mathcal{W}_{abcd}
	\\
	& + P_{ij} P_{ab} \sum_{kc} \qty( \amp{ik}{ac} \mathcal{W}_{kbcj} - \amp{i}{c} \amp{k}{a} \mel{kb}{}{cj} \VN{b} \ON{j} )
	\\
	& + P_{ij} \sum_c \amp{i}{c} \mel{ab}{}{cj} \ON{j} \VN{a} \VN{b} 
	\\
	& - P_{ab} \sum_k \amp{k}{a} \mel{kb}{}{ij} \ON{i} \ON{j} \VN{b},
\end{split}
\end{equation}
where $P$ is a permutation operator such that $P_{ij} g_{ij} = g_{ij} - g_{ji}$.
The various matrix elements from Eqs.~\eqref{eq:r1ccsd} and \eqref{eq:r2ccsd} read
\begin{subequations}
\begin{align}
\begin{split}
	\mathcal{F}_{ac} 
	& = \VN{a} F_{ac} - \frac{1}{2} \sum_k \amp{k}{a} F_{kc} 
	\\
	& + \sum_{kd} \amp{k}{d} \mel{ka}{}{dc} \VN{a} - \frac{1}{2} \sum_{kld} \tilde{\tau}_{kl}^{ad} \mel{kl}{}{cd},
\end{split}
\\
\begin{split}
	\mathcal{F}_{ki} 
	& = \ON{i} F_{ik} + \frac{1}{2} \sum_c \amp{i}{c} F_{kc}  
	\\
	& + \sum_{cl} \amp{l}{c} \mel{kl}{}{ic} \ON{i} + \frac{1}{2} \sum_{lcd} \tilde{\tau}_{il}^{cd} \mel{kl}{}{cd},
\end{split}
\\
	\mathcal{F}_{kc} & = F_{kc} + \sum_{ld} \amp{l}{d} \mel{kl}{}{cd},
\end{align}
\end{subequations}
and
\begin{subequations}
\begin{align}
\begin{split}
	\mathcal{W}_{klij} 
	& = \ON{i} \ON{j} \mel{kl}{}{ij}
	\\
	& + P_{ij} \sum_c \amp{j}{c} \mel{kl}{}{ic} \ON{i} + \frac{1}{4} \sum_{cd} \tau_{ij}^{cd} \mel{kl}{}{cd},
\end{split}
\\
\begin{split}
	\mathcal{W}_{abcd} 
	& = \VN{a} \VN{b} \mel{ab}{}{cd}
	\\
	& - P_{ab} \sum_k \amp{k}{b} \mel{ak}{}{cd} \VN{a} + \frac{1}{4} \sum_{kl} \tau_{kl}^{ab} \mel{kl}{}{cd},
\end{split}
\\
\begin{split}
	\mathcal{W}_{kbcj} 
	& = \VN{b} \ON{j} \mel{kb}{}{cj}
	\\
	& + \sum_d \amp{j}{d} \mel{kb}{}{cd} \VN{b} - \sum_{l} \amp{l}{b} \mel{kl}{}{cj} \ON{j}
	 \\
	& - \sum_{ld} \qty( \frac{\amp{jl}{db}}{2} + \amp{j}{d} \amp{l}{b} ) \mel{kl}{}{cd},
\end{split}
\end{align}
\end{subequations}
where we have defined the intermediate quantities 
\begin{subequations}
\begin{align}
	\tau_{ij}^{ab} 
	& = \amp{ij}{ab} + \amp{i}{a} \amp{j}{b} - \amp{i}{b} \amp{j}{a},
	\\
	\tilde{\tau}_{ij}^{ab} 
	& = \amp{ij}{ab} + \frac{1}{2} \qty( \amp{i}{a} \amp{j}{b} - \amp{i}{b} \amp{j}{a} ).
\end{align}
\end{subequations}
The element
\begin{equation}
	F_{pq} = \frac{1}{\ON{p}} \int \frac{  \delta \qty( E^{(0)} + E_\pair^{\DNO} )}{\delta \NSO{p} (\bx) } \NSO{q} (\bx) d\bx
\end{equation}
denotes a generalized $\DNO$ Fock matrix element, where $\NSO{p}(\bx)$ is a natural spin-orbital.  The antisymmetrized electron repulsion integrals are given by $\mel{pq}{}{rs} = \braket{pq}{rs} - \braket{pq}{sr}$ with
\begin{equation}
	\braket{pq}{rs} = \int \frac{\NSO{p}(\bx_1)\NSO{q}(\bx_2)\NSO{r}(\bx_1)\NSO{s}(\bx_2)}{r_{12}} d\bx_1 d\bx_2.
\end{equation}

Besides introducing occupancies and vacancies, and expanding the range of the sums over spin-orbitals, one more modification is applied to both the $\res{i}{a}$ and $\res{ij}{ab}$ equations. The leading term of each residual equation is multiplied by a damping factor ($\damps{i}{a}$ for $\res{i}{a}$ and $\damp{ij}{ab}$ for $\res{ij}{ab}$), which are defined in Appendix \ref{app:damp}.  \josh{The damping factors are derived by considering the difference between the zeroth-order 2-RDM, $\rRRDM^{(0)}$, and the pair-corrected and statically correlated $\DNO$ 2-RDM, $\rRRDM^{\DNO}$.  The effects of the pair correction are derived for each occupied-occupied, and virtual-virtual, spin-orbital pair and the effects of static correlation and the high-spin correction are derived by considering each statically correlated electron pair.}

The CCSD energy expression is unmodified, with the exception of the range of summation,
\begin{equation} 
	\label{eq:eccsd}
	E^{\DNO}_\text{CCSD} = \sum_{ia} \amp{i}{a} F_{ia} + \frac{1}{2} \sum_{ijab} \qty( \frac{ \amp{ij}{ab}}{2} - \amp{i}{a} \amp{j}{b}  ) \mel{ij}{}{ab}.
\end{equation}

\subsection{MP2 for $\DNO$}
\label{subsec:mp2}
An equation for the MP2 amplitudes is derived in a manner analogous to CCSD, where the two instructions of White and Chan [see Subsec.~\ref{subsec:ccsd}] are applied to the usual non-canonical MP2 residual equation,
\begin{equation} 
\label{eq:r2mp2}
\begin{split} 
	\res{ij}{ab}(\text{MP2}) 
	& = \ON{i} \ON{j} \VN{a} \VN{b} \damp{ij}{ab} \mel{ij}{}{ab} 
	\\ 
	& + \sum_c \qty( \VN{b} \amp{ij}{ac} F_{bc} + \VN{a} \amp{ij}{cb} F_{ac} )
	\\
	& - \sum_k \qty( \ON{j} \amp{ik}{ab} F_{jk} + \ON{i} \amp{kj}{ab} F_{ik} ),
\end{split}
\end{equation}
where, in addition to the introduction of occupancies and vacancies, and the expanded range of summation, the same damping factor (defined in  Appendix \ref{app:damp}) applied to the CCSD $\res{ij}{ab}$ equation [see Eq.~\eqref{eq:r2ccsd}] is applied here. Also like CCSD, the MP2 energy expression remains the same with the exception of the expanded range of summation, \textit{i.e.},
\begin{equation} 
\label{eq:emp2}
	E^{\DNO}_\text{MP2} = \frac{1}{4} \sum_{ijab} \amp{ij}{ab} \mel{ij}{}{ab}.
\end{equation}
Note that because the non-canonical formulation of MP2\cite{Pulay1986} is employed, Eqs.~\eqref{eq:r2mp2} and \eqref{eq:emp2} do not involve single excitations.

\section{Method}
\label{sec:method}

\subsection{FCI reference}
\label{subsec:exFCI}
Benchmark potential energy curves were obtained using a determinant-driven selected configuration interaction (sCI) method known as CIPSI (Configuration Interaction using a Perturbative Selection made Iteratively)\cite{Huron1973,Giner2013,Giner2015} in which the energies are extrapolated to the full configuration interaction (FCI) result using multireference perturbation theory.\cite{Garniron2017,Loos2018,qp2}  The all-electron extrapolated-FCI (exFCI) calculations were performed using Quantum Package 2.0.\cite{qp2} All benchmark and $\DNO$ calculations were performed using the cc-pVTZ/f basis set.\cite{Dunning1989,Prascher2011,Feller1996,Schuchardt2007}

\subsection{$\DNO$}
\label{subsec:DNOmethod}

All $\DNO$ and subsequent MP2 and CCSD calculations were performed using the MUNgauss quantum chemistry program.\cite{MUNgauss}  Optimization of the $\qty{ \NO{m} }$ and the $\qty{ \Delta_{me} }$ was performed according to the previously established algorithm.\cite{Piris2009,NOF1}  Restricted Hartree-Fock orbitals serve as the initial guess NOs, which are then optimized via iterative diagonalization of a pseudo-Fock matrix.  The $\qty{ \Delta_{me} }$ are optimized using a Newton-Raphson algorithm.  For the current study, the number of active occupied orbitals was chosen manually to be the number of bonds in the diatomic. An equivalent number of virtual orbitals were chosen to be active.

\subsection{MP2 and CCSD}
\label{subsec:MP2CCSDmethod}
Both the MP2 and CCSD algorithms are implemented in the spin-orbital basis.  The residual equations are solved using an iterative Newton-Raphson (MP2), or approximate Newton-Raphson (CCSD), approach. Amplitude updates are calculated via
\begin{subequations}
\begin{align}
	\amp{i}{a} & \leftarrow \amp{i}{a} + \frac{\res{i}{a}}{\ON{i} F_{ii} - \VN{a} F_{aa}},
	\\
	\amp{ij}{ab} & \leftarrow \amp{ij}{ab} + \frac{\res{ij}{ab}}{\ON{i} F_{ii} + \ON{j} F_{jj} - \VN{a} F_{aa} - \VN{b} F_{bb}},
\end{align}
\end{subequations}
where the initial $\amp{ij}{ab}$ amplitudes are set to the MP2 values and $\amp{i}{a} = 0$. 
To avoid numerical instabilities, residuals, $\res{i}{a}$ and $\res{ij}{ab}$, are considered to be zero if the leading term [see Eqs.~\eqref{eq:r1ccsd} and \eqref{eq:r2ccsd} for CCSD, and Eq.~\eqref{eq:r2mp2} for MP2] is below a specific threshold $\tau$,
\begin{subequations}
\begin{align}
	\res{i}{a} & = 0, &\text{ if } & \ON{i} \VN{a} \damp{i}{a} F_{ia} < \tau,
	\\
	\res{ij}{ab} & = 0, &\text{ if } & \ON{i} \ON{j} \VN{a} \VN{b} \damp{ij}{ab} \mel{ij}{}{ab} < \tau.
\end{align}
\end{subequations}
Here $\tau$ is set to machine precision.
The iterative optimization of the amplitudes is accelerated using a direct inversion of iterative subspace (DIIS) algorithm\cite{DIIS} to extrapolate from amplitudes of previous steps. \cite{Scuseria1986} A maximum number of ten sets of amplitudes from previous steps were kept for extrapolation. In the case of the CCSD iterations, the $\amp{i}{a}$ and $\amp{ij}{ab}$ were combined and extrapolated together.  Convergence was assumed when the absolute value of the largest residual element was less than $10^{-7}$.

\section{Results}
\label{sec:results}
\begin{figure}
	\begin{center}
	\includegraphics[width=0.8\textwidth]{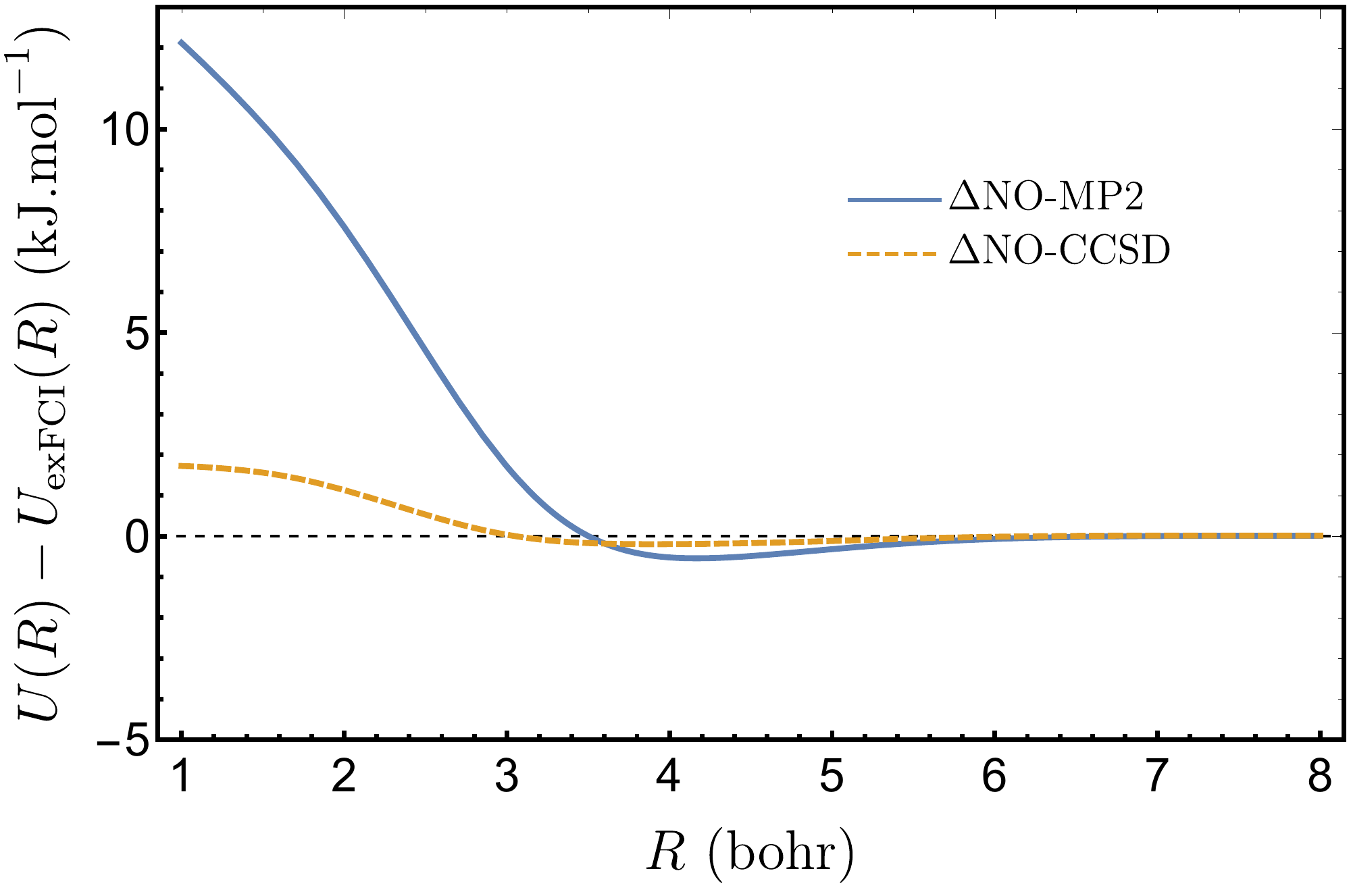}
	\caption{\label{fig:UerrorH2} Error (in kJ.mol$^{-1}$) in \ce{H2} potential energy curves compared to exFCI. The error in $U_\text{MP2}(R)$ is 17 kJ.mol$^{-1}$ at $R=1$ and continues to grow with increasing $R$. For \ce{H2}, $U_\text{CCSD}(R) = U_\text{exFCI}(R)$ and therefore the CCSD error is zero for all $R$.}
	\end{center}
\end{figure}

The error in the $\DNO$-MP2 and $\DNO$-CCSD potential energy curves, $U(R)$, for \ce{H2} compared to exFCI is presented in Fig.~\ref{fig:UerrorH2}.
The potential energy curve is calculated as 
\begin{equation}
	U(R) = E(R) + V_\text{nn}(R),
\end{equation}
where the nuclear repulsion energy, $V_\text{nn}(R)$, is added to the electronic energy obtained from $\DNO$-MP2, $\DNO$-CCSD, or exFCI.  For \ce{H2}, the exFCI result is equivalent to regular FCI, and hence, the exact result for the given basis set.  Also, for two electrons, CCSD is equivalent to FCI and therefore any error in $U_{\Delta\text{NO-CCSD}}(R)$ is due to the manner in which the $\DNO$ static correlation energy is blended with the CCSD dynamic correlation energy.  This leads to a maximum error of $1.7$ kJ.mol$^{-1}$ at the beginning of the examined range, $R=1$.  There is also a slight overestimation of the total correlation energy at stretched bond lengths, with a maximum deviation of $-0.2$ kJ.mol$^{-1}$ at $R=3.89$.  In the case of $\DNO$-MP2, the error at small $R$ is much larger.  This can be attributed to the fact that, as $R\to 0$, the correlation energy approaches that of \ce{He}, for which the MP2 correlation energy differs from the FCI correlation energy by $15.5$ kJ.mol$^{-1}$.  As $R$ increases the error in $U_{\Delta\text{NO-MP2}}(R)$ decreases, also with a slight overestimation of correlation energy ($-0.5$ kJ.mol$^{-1}$ at $R=4.17$) at stretched bond lengths.  For both methods, the damping factors ensure that, as $R\to \infty$, the dynamic correlation energy vanishes, along with the error in $U(R)$. 

Equilibrium bond lengths and dissociation energies predicted by $\DNO$-MP2 and $\DNO$-CCSD for a selection of diatomics, are compared to $\DNO$ (no dynamic correlation), NOF-MP2\cite{NOFMP2b}, MP2, CCSD and exFCI values in Table 
\ref{tab:ReDe}.
\begin{turnpage}
\begin{table*}
	\caption{\label{tab:ReDe} Calculated equilibrium bond lengths $R_\text{e}$ and dissociation energies $D_\text{e}$ for a selection of diatomics.}
	\begin{ruledtabular}
	\begin{tabular}{lccccccccccccccc}
				& \multicolumn{7}{c}{$R_\text{e}$ (bohr)} & \multicolumn{7}{c}{$D_\text{e}$ (kJ.mol$^{-1}$)} 
		\\ 
		\cline{2-8} \cline{10-16}
		Molecule & exFCI & MP2 & CCSD & $\DNO$\footnotemark[2] & $\DNO$-MP2 & $\DNO$-CCSD & NOF-MP2\footnotemark[3] & & exFCI & MP2\footnotemark[1] & CCSD\footnotemark[1] & $\DNO$\footnotemark[2] & $\DNO$-MP2 & $\DNO$-CCSD & NOF-MP2\footnotemark[3] \\
		\hline
\ce{H2}	& 1.405 & 1.392 & 1.405 & 1.428 & 1.408 & 1.405 &       & & 454 &     & 454 & 399 & 443 & 452 &     \\
\ce{LiH}& 3.028 & 3.019 & 3.027 & 3.019 & 3.044 & 3.030 &       & & 236 & 343 & 238 & 185 & 223 & 233 &     \\
\ce{HF}	& 1.729 & 1.731 & 1.725 & 1.735 & 1.741 & 1.741 & 1.731 & & 576 & 667 & 641 & 470 & 603 & 606 & 590 \\
\ce{LiF}& 2.981 & 2.986 & 2.976 & 2.946 & 2.991 & 2.979 & 2.984 & & 549 & 612 & 610 & 424 & 593 & 582 & 590 \\
\ce{F2}	& 2.692 & 2.655 & 2.649 & 2.779 & 2.623 & 2.634 & 2.612 & & 143 &     & 268 & \hphantom{0}67    & 138 & 148 & 192 \\
\ce{N2}	& 2.083 & 2.102 & 2.069 & 2.071 & 2.084 & 2.077 & 2.075 & & 880 &     & 780 & 712 & 856 & 891 & 965
	\end{tabular}
	\end{ruledtabular}
	\footnotetext[1]{$D_e$ for potential energy curves with singularities are not reported.}
	\footnotetext[2]{From potential energy curve where $E^{\DNO}_\text{dyn} = 0$}
	\footnotetext[3]{Orbital-invariant formulation of NOF-MP2 (NOF-OIMP2/cc-pVTZ) from Piris.\cite{NOFMP2b}}
\end{table*}
\end{turnpage}
As expected from Fig.~\ref{fig:UerrorH2}, the \ce{H2} $R_e$ and $D_e$ values predicted by $\DNO$-MP2 and $\DNO$-CCSD are very close to the exFCI values.  The underestimation of $D_e$ by $\DNO$-MP2 is attributable to the lack of dynamic correlation at small to intermediate $R$. Removal of all of the dynamic correlation, by using $\DNO$, results in a much larger underestimation of $D_e$ (by $55$ kJ.mol$^{-1}$).   No MP2 $D_e$ value is reported due to the well-known divergence of the potential energy curve to $-\infty$ as $R$ increases.  The divergence is due to the degeneracy of the $\sigma$-bonding and $\sigma^*$-antibonding orbitals of \ce{H2} as $R\to\infty$, and is completely removed in the $\DNO$-MP2 treatment.

\begin{figure}
	\begin{center}
	\includegraphics[width=0.8\textwidth]{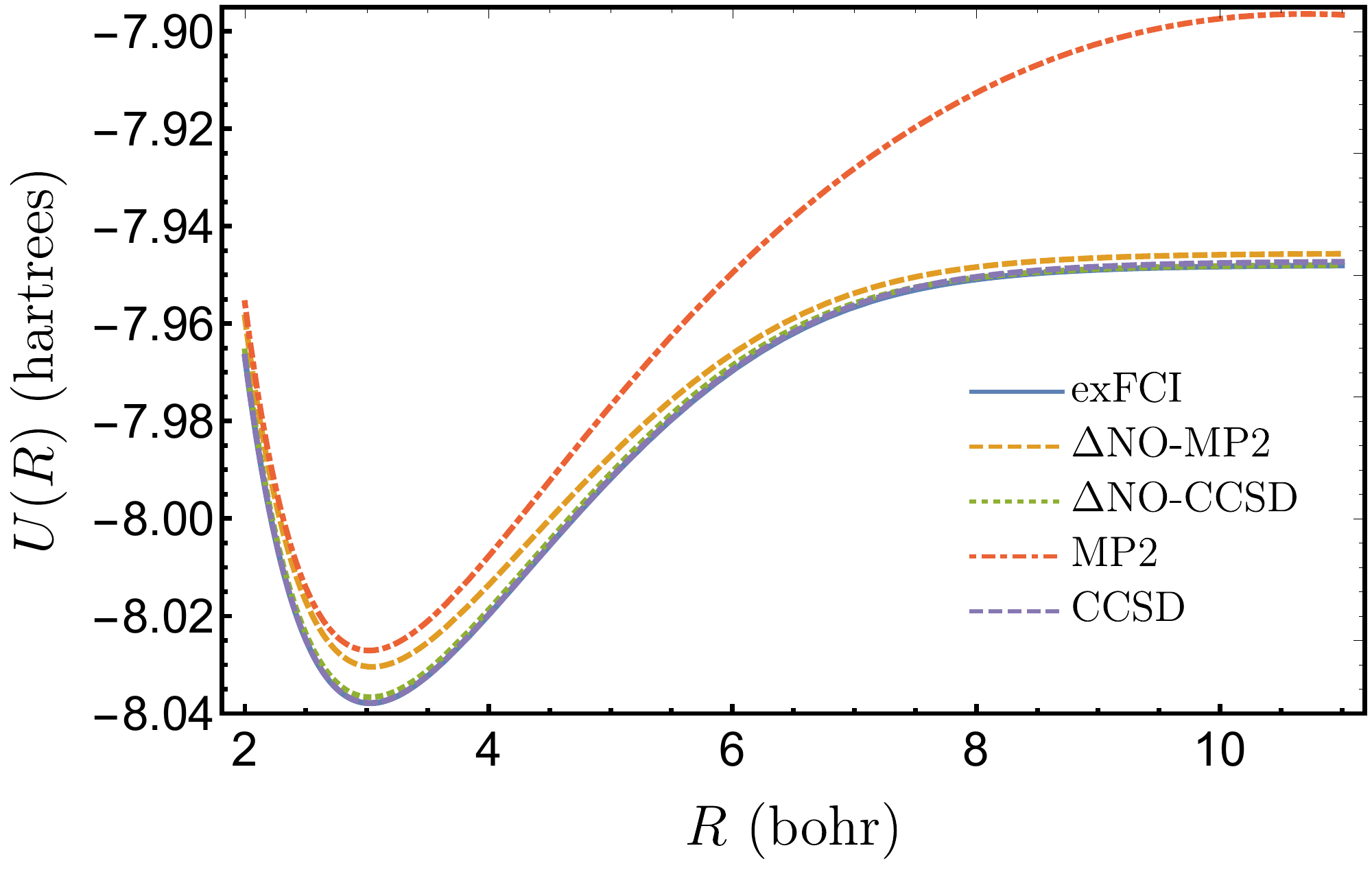}	
	\caption{\label{fig:LiHPES} Calculated \ce{LiH} potential energy curves.}
	\end{center}
\end{figure}

The $\DNO$-MP2, $\DNO$-CCSD, MP2, CCSD and exFCI curves for \ce{LiH} are presented in Fig.~\ref{fig:LiHPES}.
Both $\DNO$-MP2 and $\DNO$-CCSD provide an accurate model of \ce{LiH} dissociation. Most of the error in $U_{\Delta\text{NO-MP2}}(R)$ occurs near equilibrium, deviating from $U_{\text{exFCI}}(R)$ by $20$ kJ.mol$^{-1}$ at $R=3.028$. Whereas, $U_{\Delta\text{NO-CCSD}}(R)$ only deviates by $3$ kJ.mol$^{-1}$ at equilibrium, and $0.2$ kJ.mol$^{-1}$ near dissociation, $R=11$.  This means both the static correlation of the \ce{LiH} bond and dynamic correlation of the electrons on \ce{Li} are effectively captured by $\DNO$-CCSD. Inclusion of only static correlation, via $\DNO$, leads to a reasonable prediction of $R_e$ ($3.019$) but $D_e$ is underestimated by $51$ kJ.mol$^{-1}$. 

\begin{figure}
	\begin{center}
	\includegraphics[width=0.8\textwidth]{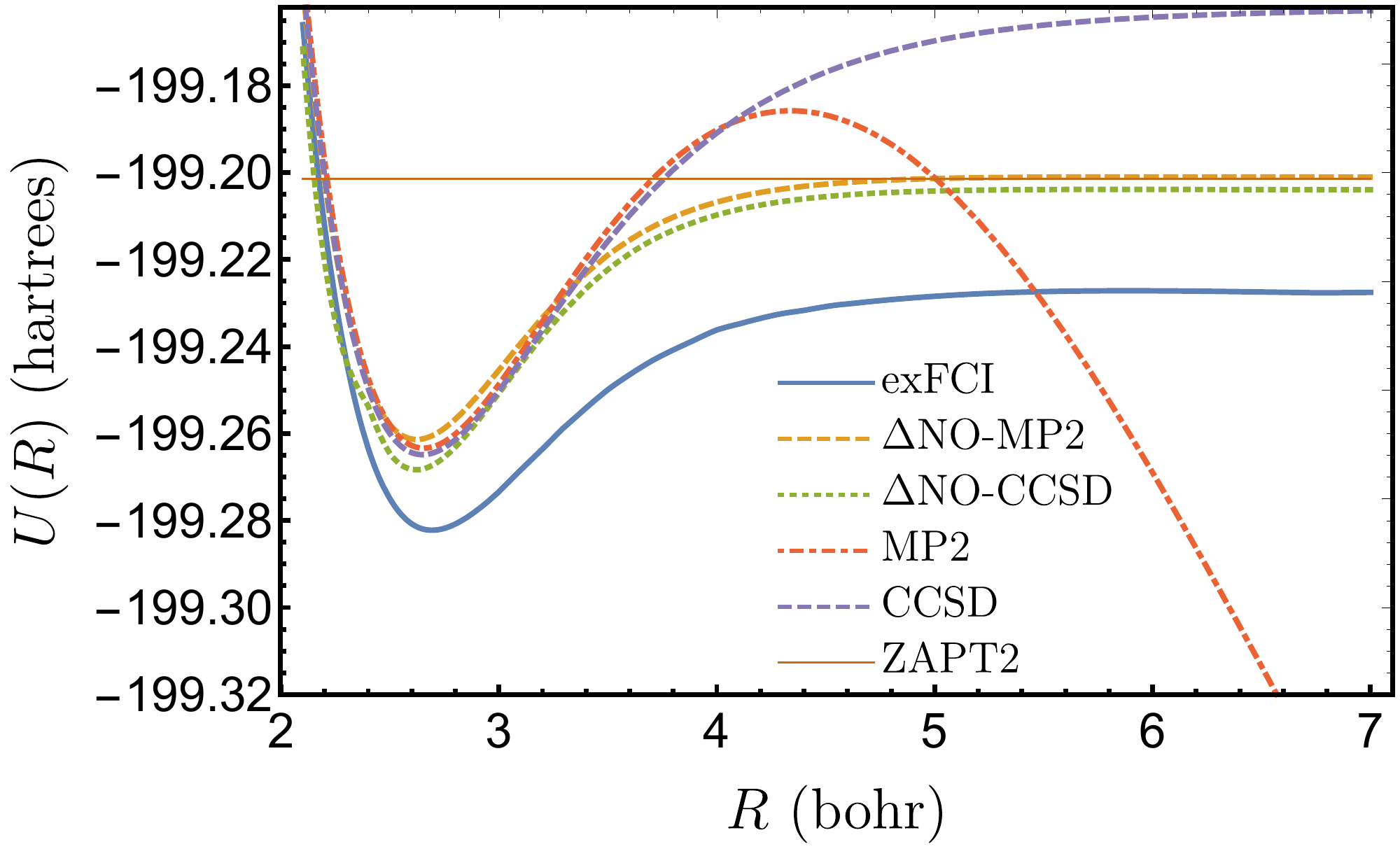}
	\caption{\label{fig:F2PES} Calculated \ce{F2} potential energy curves. The ZAPT2 energy of two separate \ce{F} atoms is included for comparison.}
	\end{center}
\end{figure}

\begin{figure}
	\begin{center}
	\includegraphics[width=0.8\textwidth]{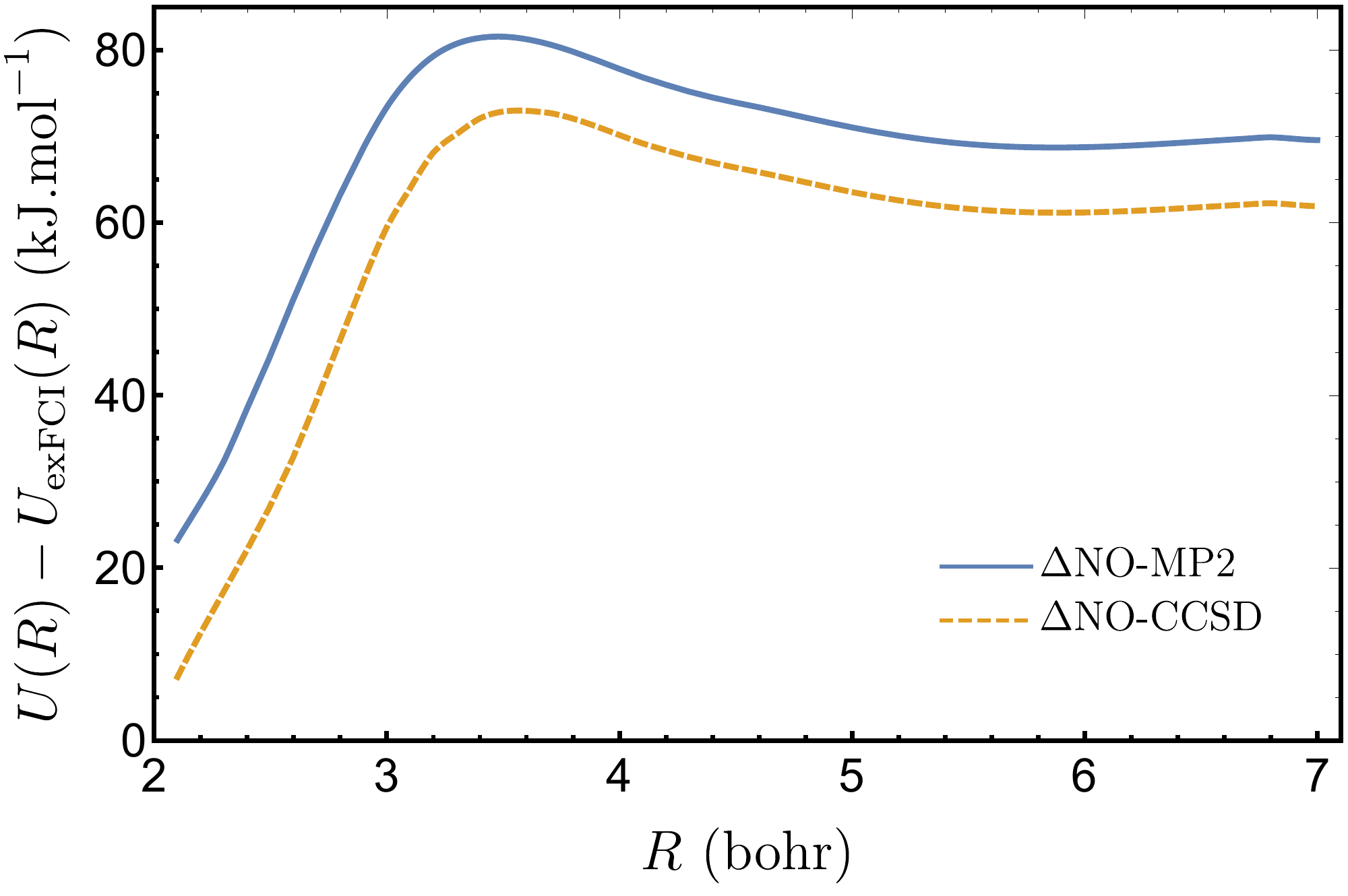}
	\caption{\label{fig:F2error} Error (in kJ.mol$^{-1}$) in \ce{F2} potential energy curves compared to exFCI.}
	\end{center}
\end{figure}

The potential energy curves for \ce{F2} are shown in Fig.~\ref{fig:F2PES}\josh{, and the error in $U_{\Delta\text{NO-MP2}}(R)$ and $U_{\Delta\text{NO-CCSD}}(R)$ compared to $U_{\text{exFCI}}(R)$ is shown in Fig.~\ref{fig:F2error}.}
Similar to other post-Hartree-Fock correlation methods, $\DNO$-MP2 and $\DNO$-CCSD struggle to capture the dynamic correlation in \ce{F2} near equilibrium and in the separated \ce{F} atoms.\cite{Giner2015} This is evident in the large separation between the exFCI curve and all the others.  In the case of $\DNO$-MP2 and $\DNO$-CCSD, the lack of dynamic correlation is relatively consistent and therefore the predicted $D_e$ values are reasonable for both, with $\DNO$-MP2 differing from exFCI by $-5$ kJ.mol$^{-1}$ and $\DNO$-CCSD differing by $+5$ kJ.mol$^{-1}$.  The lack of dynamic correlation has a more significant effect on the predicted $R_e$ values, which differ by $-0.07$ for $\DNO$-MP2 and $-0.06$ for $\DNO$-CCSD.  \josh{This is unsurprising when considering that the error in both $U_{\Delta\text{NO-MP2}}(R)$ and $U_{\Delta\text{NO-CCSD}}(R)$ varies the most around $R_e$ (Fig.~\ref{fig:F2error}).}  The predicted $R_e$ values from MP2 and CCSD also deviate negatively from the exFCI $R_e$, but the deviation is smaller, approximately $-0.04$. If dynamic correlation is completely neglected ($\DNO$), $R_e$ is significantly overestimated ($+0.09$), and the estimated $D_e$ is exceptionally small, $67$ kJ.mol$^{-1}$.  

For a given molecule, the $\DNO$ energy without dynamic correlation energy (simply referred to as $\DNO$ in Table \ref{tab:ReDe}), $E^{\DNO}_\text{no-dyn} = E^{(0)} + E^{\DNO}_\pair + E^{\DNO}_\stat + E^{\DNO}_\HSC$, is equivalent to the sum of restricted open-shell Hartree-Fock (ROHF) energies at the bond dissociation limit,
\begin{equation} \label{DNOandROHF}
	\lim_{R\to\infty} E^{\DNO}_\text{no-dyn} [ \ce{A \overset{R}{\bond{...}} B} ] = E_\text{ROHF}[\ce{A}] + E_\text{ROHF}[\ce{B}].
\end{equation}
Therefore, the quality of the $\DNO$-MP2 treatment near the bond dissociation limit can be assessed through comparison of the $\DNO$-MP2 energy to the ROHF energy plus the $z$-averaged second-order perturbation energy (ZAPT2)\cite{Lee1993} of the two separated fragments. The ZAPT2 energy of two \ce{F} atoms is plotted in Fig.~\ref{fig:F2PES}, where it is seen that the $\DNO$-MP2 energy is $1.0$ kJ.mol$^{-1}$ higher.  This confirms that $\DNO$-MP2 is correctly capturing and partitioning the static and dynamic correlation energy of \ce{F2}. This is in sharp contrast to CCSD which drastically overestimates $D_e$, or MP2 which diverges due to orbital degeneracy.  It is clear that, contrary to conventional single-reference methods like MP2 and CCSD, the hybrid $\DNO$-MP2 and $\DNO$-CCSD methods proposed here are able to accurately model strongly correlated systems.

Similar to \ce{F2}, the dynamic correlation of the \ce{F} atom in \ce{HF} and \ce{LiF} is not sufficiently captured by $\DNO$-MP2 or $\DNO$-CCSD.  This leads to overestimation of $D_e$ compared to exFCI (see Table \ref{tab:ReDe}).  However, the lack of static correlation in MP2 and CCSD leads to even larger overestimation of $D_e$.

\begin{figure}
	\begin{center}
	\includegraphics[width=0.8\textwidth]{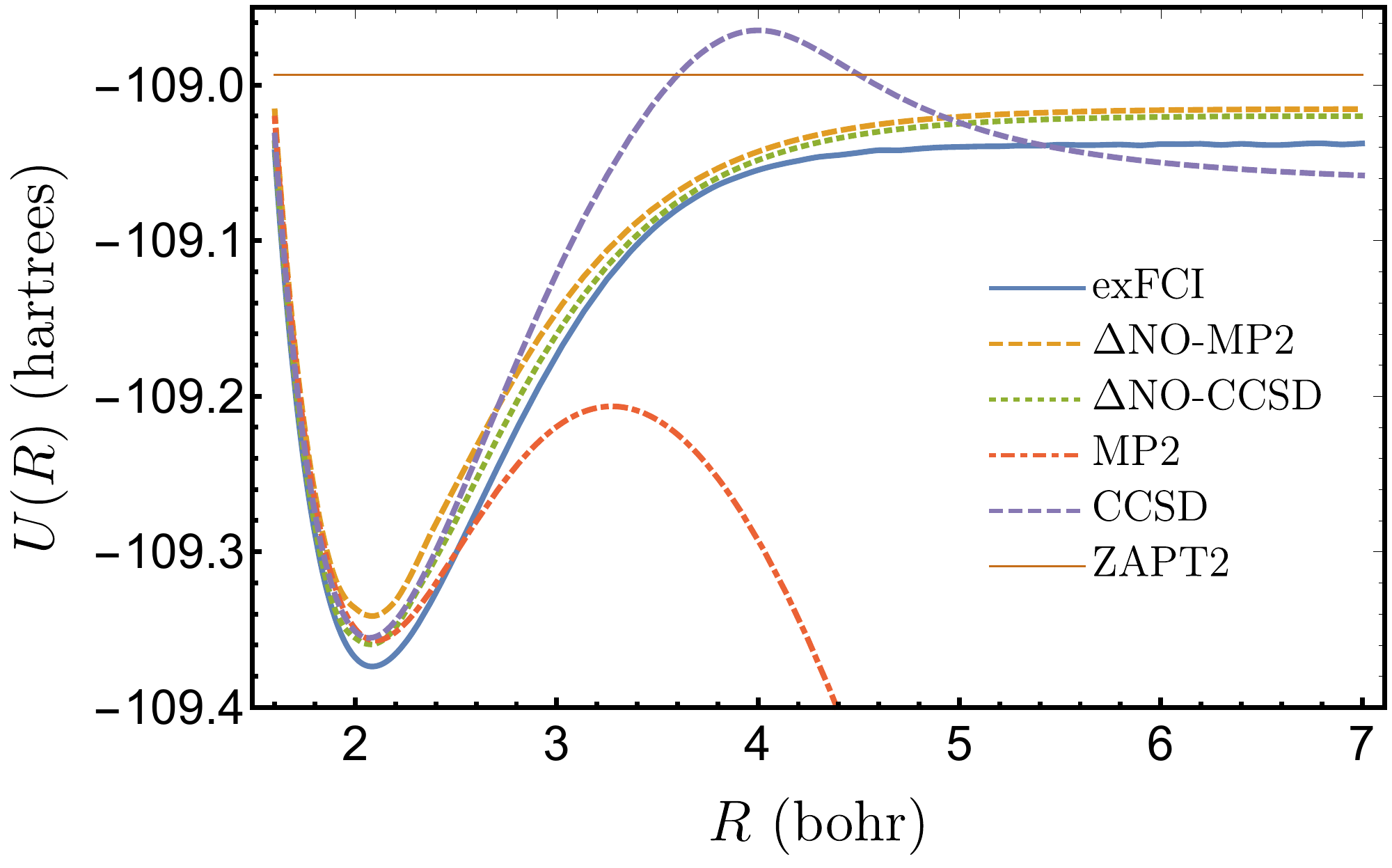}
	\caption{\label{fig:N2PES} Calculated \ce{N2} potential energy curves. The ZAPT2 energy of two separate \ce{N} atoms is included for comparison.}
	\end{center}
\end{figure}

\begin{figure}
	\begin{center}
	\includegraphics[width=0.8\textwidth]{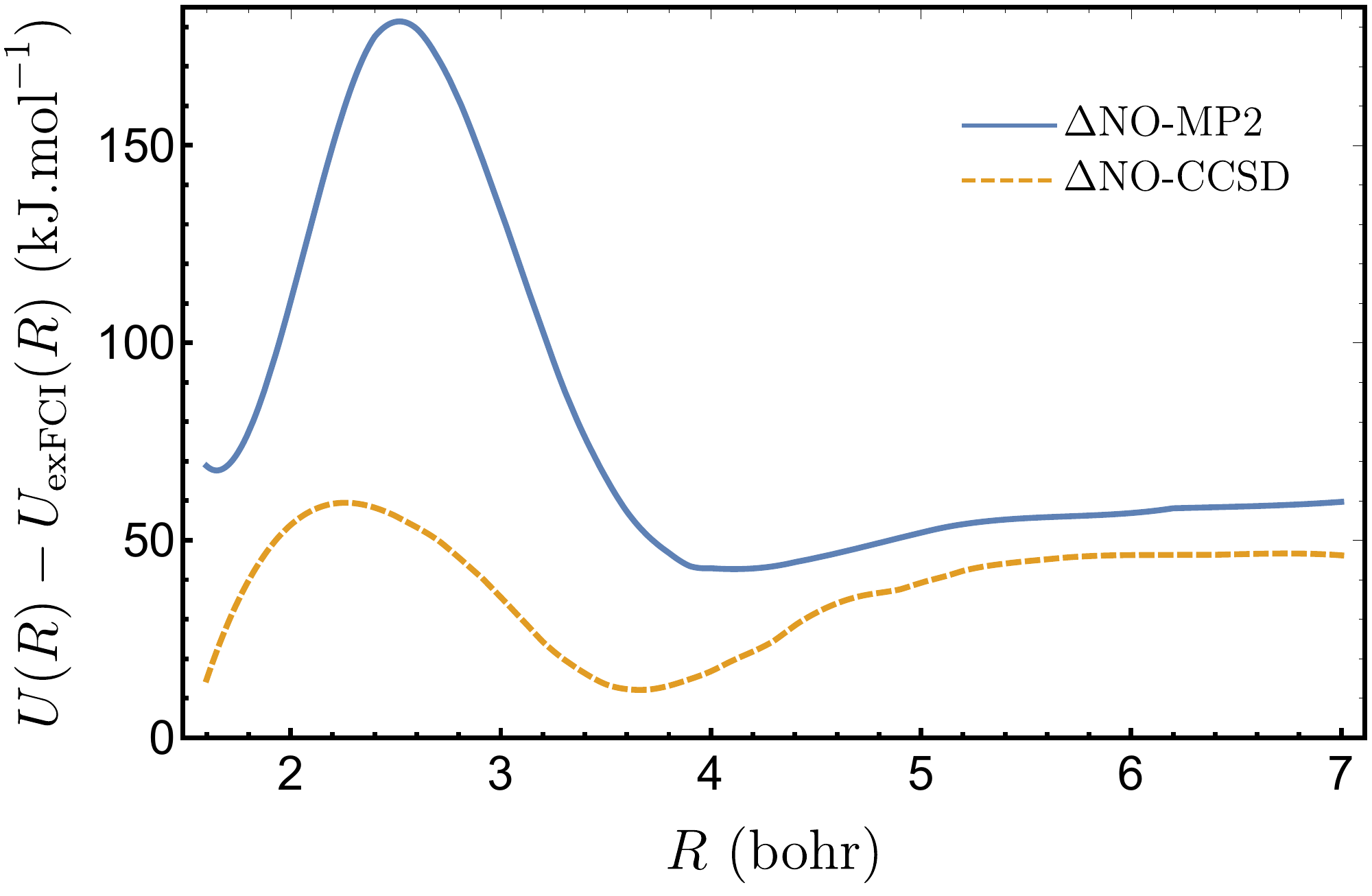}
	\caption{\label{fig:N2error} Error (in kJ.mol$^{-1}$) in \ce{N2} potential energy curves compared to exFCI.}
	\end{center}
\end{figure}

In Fig.~\ref{fig:N2PES}, the $\DNO$-MP2 and $\DNO$-CCSD potential energy curves for \ce{N2} are compared to that of MP2, CCSD and exFCI. \josh{The error in $U_{\Delta\text{NO-MP2}}(R)$ and $U_{\Delta\text{NO-CCSD}}(R)$ compared to $U_{\text{exFCI}}(R)$ is also shown in Fig.~\ref{fig:N2error}.}
Like \ce{F2}, $\DNO$-MP2 overestimates $D_e$ while $\DNO$-CCSD underestimates it.  Albeit, the amount by which the $\DNO$ methods are in error is somewhat greater, $-24$ kJ.mol$^{-1}$ for $\DNO$-MP2 and $+11$ kJ.mol$^{-1}$ for $\DNO$-CCSD.  Significantly more dynamic correlation is captured by $\DNO$-CCSD near equilibrium compared to $\DNO$-MP2.  At the exFCI equilibrium bond length, $R_e=2.083$, $U_{\Delta\text{NO-CCSD}}(R)$ is $38$ kJ.mol$^{-1}$ above $U_{\text{exFCI}}(R)$, whereas $U_{\Delta\text{NO-MP2}}(R)$ is $85$ kJ.mol$^{-1}$ above. Both predicted equilibrium bond lengths are in good agreement with the exFCI values, particularly the $\DNO$-CCSD value of $R_e =2.084$.  At dissociation, both methods underestimate the dynamic correlation by similar amounts, $60$ kJ.mol$^{-1}$ for $\DNO$-MP2 and $53$ kJ.mol$^{-1}$ for $\DNO$-CCSD.  Interestingly, the $\DNO$-MP2 energy at dissociation is $58$ kJ.mol$^{-1}$ lower than the ZAPT2 result for two separate \ce{N} atoms.  Analysis of the components of the ZAPT2 and $\DNO$-MP2 correlation energies reveals it is the correlation between the statically correlated electrons (\textit{i.e.},~unpaired electrons) which is responsible for this difference.  \josh{This discrepency, for \ce{F2} and \ce{N2}, suggests that both $\DNO$-MP2 and $\DNO$-CCSD are not size-consistent. However, further analysis is required to reveal the origin of, and fully understand, the discrepancy.}

For the small collection of molecules studied, the quality of $\DNO$-MP2 and $\DNO$-CCSD improves, compared to the single-reference MP2 and CCSD, as the complexity of the system increases. Expectedly, as the amount of static correlation increases ({\it i.e.},~small $R$ to large $R$, or single bond to triple bond) the $\DNO$ methods become significantly superior.

\FloatBarrier
\section{Conclusions}
\label{sec:conc}

Combining multireference methods for static correlation with post-Hartree-Fock methods for dynamic correlation is a common approach to modeling complex electronic systems.  Despite the noted success of CFT methods in modeling systems with multireference character, there is only one example of using such a method in combination with post-Hartree-Fock correlation, which is NOF-MP2. In this work, a CFT method, $\DNO$, is combined with both MP2 and CCSD in a fashion completely analogous to each other.  This is achieved by incorporating occupancies and vacancies, and expanded domains for occupied and virtual orbitals, according to guidelines used to derive FT-CCSD.  Additionally, the MP2 and CCSD correlation energies are combined with $\DNO$ by inserting $\Delta$-dependent damping factors into the residual equations. The damping factors are defined by considering the description of statically correlated electron pairs by the $\DNO$ 2-RDM, particularly the spin-orbitals they simultaneously occupy (and vacate) as static correlation becomes appreciable.

For the six diatomics studied, both $\DNO$-MP2 and $\DNO$-CCSD predict reasonable bond lengths and dissociation energies compared to the benchmark exFCI values.  The error in the descriptions of \ce{HF}, \ce{LiF} and \ce{F2} is largely due to the inability of MP2, or CCSD, to account for all of the dynamic correlation amongst the electrons of \ce{F}.  The larger error in $D_e$ values predicted for \ce{N2} is likely due to the fact that three bonds are being broken compared to one in the other diatomics. However, the discrepancy between the $\DNO$-MP2 energy for dissociated \ce{N2} and the ZAPT2 energy for two \ce{N} atoms deserves attention. That, in combination with the discrepancy between the CCSD and $\DNO$-CCSD curves for \ce{H2}, suggests that further analysis, of the modified MP2 and CCSD equations in particular, could lead to a more seamless fusion of $\DNO$ and post-Hartree-Fock methods.

In addition to providing an alternative treatment of multireference systems, the $\DNO$-MP2 and $\DNO$-CCSD methods offer insight into static and dynamic correlation and the balance between the two.  Although most definitions of static correlation make use of the concept of degeneracy or near-degeneracy, the methods presented here are free from any such arguments.  The damping factors are based on the simultaneous occupancy (or vacancy) of active spin-orbitals.  The damping factors modify the MP2 and CCSD residual equations according to how the $\DNO$ static correlation influences the 2-RDM. Such concepts are relatively easy to grasp in the limit of complete static correlation, and provide a useful ``physical" picture of a multireference system.

\begin{acknowledgements}
JWH thanks the Natural Sciences and Engineering Research Council of Canada (NSERC) for a Discovery Grant, Compute/Calcul Canada for computing resources, the Universit\'e Paul Sabatier (Toulouse, France) for a visiting fellowship, and the Discovery Institute for Computation and Synthesis for useful consultations.  The authors also thank Anthony Scemama for helpful discussions.
\end{acknowledgements}

\appendix

\section{Amplitude damping}
\label{app:damp}
The MP2 and CCSD equations adopted from the finite-temperature versions [see Subsecs.~\ref{subsec:ccsd} and \ref{subsec:mp2}], are modified for use with the $\DNO$ method by incorporating a single excitation, $\damps{i}{a}$, and a double excitation, $\damp{ij}{ab}$, damping factor. The single excitation damping factor included in the CCSD $\res{i}{a}$ equation takes a rather simple form,
\begin{equation}
	\damps{i}{a} = 
		\begin{cases}
		0, &\text{ if } \NO{i} \in \Act,
		\\
		1, & \text{ otherwise}.
		\end{cases} 
\end{equation}
In other words, all single excitations from the active $\DNO$ orbitals are turned off. This arises from the assumption that single excitations are responsible for orbital relaxation, \cite{Scuseria1987,Sherrill1998} and that the most significant part of the active orbital relaxation (due to static correlation) is obtained via the $\Delta$NO orbital optimization.

\josh{
For double excitations, the damping factor,
\begin{equation}
	\damp{ij}{ab} = \alpha_{ij} \alpha^{ab} \beta_{ij} \beta^{ab} \beta_i^a \beta_j^b \beta_i^b \beta_j^a,
\end{equation}
is comprised of pair correction, $\alpha_{ij}\alpha^{ab}$, and static correlation and high-spin correction, $\beta_{ij} \beta^{ab} \beta_i^a \beta_j^b \beta_i^b \beta_j^a$, contributions. 
Terms are defined for each possible occupied-virtual pairing, $\beta_i^a \beta_j^b \beta_i^b \beta_j^a$, to maintain the symmetry of the amplitudes,  $\amp{ij}{ab} = -\amp{ji}{ab} = - \amp{ij}{ba} = \amp{ji}{ba}$.
}

\josh{
In the double excitation residual equations for both MP2 [Eq.~\eqref{eq:r2mp2}] and CCSD [Eq.~\eqref{eq:r2ccsd}] the damping factor is applied to the leading term which includes a product of the orbital occupancies and vacancies, $\ON{i} \ON{j} \VN{a} \VN{b}$.  
The pair-correction contribution to the damping factor, $\alpha_{ij}$ and $\alpha^{ab}$, correct the occupancy-occupancy and vacancy-vacancy products, respectively. 
The occupancy-occupancy term is defined as,
\begin{align}
	\alpha_{pq} & = \frac{\left( \rRRDM^{(0),\up\up} \right)_{pqpq} + \left( \rRRDM^{\DNO,\up\up}_{\pair} \right)_{pqpq}}{\left( \rRRDM^{(0),\up\up} \right)_{pqpq}}, 
	&
	\alpha_{p\bar{q}} & = \frac{\left( \rRRDM^{(0),\up\dw} \right)_{pqpq} + \left( \rRRDM^{\DNO,\up\dw}_{\pair} \right)_{pqpq}}{\left( \rRRDM^{(0),\up\dw} \right)_{pqpq}}.
\end{align}
As in the MP2 and CCSD equations, the indices of the damping factors refer to spin-orbitals.  
In the $\DNO$ method, terms are labelled according to spatial NOs. 
Here, the labelling ($p,q$) corresponds to spatial NOs, with spin-orbitals distinguished using an overbar for spin-down orbitals, and no overbar for spin-up orbitals.  
The vacancy-vacancy pair-correction factor, $\alpha^{ab}$, is defined in the same manner as that for occupancy-occupancy, $\alpha_{ij}$, except all occupancies ($\ON{p}$) are replaced by vacancies ($\VN{p}$), and the sign of the $\Delta$'s are reversed because they have the opposite effect on vacancies,
\begin{equation}
	\alpha^{pq} = \left. \alpha_{pq} \right|_{\substack{n \to h \\ \Delta \to -\Delta}}
\end{equation}
From these equations, expressions for the pair-correction factors for occupancy-occupancy and vacancy-vacancy pairs can be determined, for spin-orbitals from the same NO,
\begin{align}
	\alpha_{p\bar{p}} & = \frac{1}{\ON{p}}, 
	& 
	\alpha^{p\bar{p}} & = \frac{1}{\VN{p}},
\end{align}
and from different NOs,
\begin{align}
	\alpha_{pq} & = \alpha_{p\bar{q}} = \frac{(\ON{p} + \Delta_{pq})(\ON{q} - \Delta_{pq}) - \eta_{pq} }{\ON{p} \ON{q}},
	\\
	\alpha^{pq} & =\alpha^{p\bar{q}} = \frac{(\VN{p} - \Delta_{pq})(\VN{q} + \Delta_{pq}) - \eta_{pq} }{\VN{p} \VN{q}}.
\end{align}
}

\josh{
The remaining contribution to the damping factors is from the static correlation and the high-spin correction of the 2-RDM.  
Each spin-orbital pair contribution is also defined by a ratio of 2-RDM components, however, in this case the difference between the numerator and the denominator is the contribution from static correlation and the high-spin correction,
\begin{align} 
\label{eq:betamn}
	\beta_{mn} & = \frac{G^{\DNO,\up\up}_{mn}}{G^{(0),\up\up}_{mn} + \left( G^{\DNO,\up\up}_{\pair} \right)_{mn}},
	&
	\beta_{m\bar{n}}  & = \frac{G^{\DNO,\up\dw}_{mn}}{G^{(0),\up\dw}_{mn} + \left( G^{\DNO,\up\dw}_{\pair} \right)_{mn}}.
\end{align}
The above definitions only apply to active-occupied spin orbitals (denoted by $m, \bar{m}, n,$ and $\bar{n}$). 
These quantities are defined in terms of sums over 2-RDM elements,
\begin{equation}
	G^{\DNO,\up\up}_{mn} = \sum_{pq} \left( \left. \rRRDM^{\DNO,\up\up}_{pqpq} \right|_{\substack{ \ON{k} = 0\\ \Delta_{ke} = 0 \\ (k \ne m,n) }} - \left. \rRRDM^{\DNO,\up\up}_{pqpq} \right|_{\substack{ \ON{k} = 0\\ \Delta_{ke} = 0 \\ (k \ne m) }} - \left. \rRRDM^{\DNO,\up\up}_{pqpq} \right|_{\substack{ \ON{k} = 0\\ \Delta_{ke} = 0 \\ (k \ne n) }}\right),
\end{equation}
\begin{equation}
	G^{\DNO,\up\dw}_{m\bar{n}} = 
	\begin{cases} 
		\sum_{pq} \left. \rRRDM^{\DNO,\up\dw}_{ppqq} \right|_{\substack{ \ON{k} = 0\\ \Delta_{ke} = 0 \\ (k \ne m) }} 
		&, \text{if } m = n \\
		\sum_{pq} \left( \left. \rRRDM^{\DNO,\up\dw}_{pqpq} \right|_{\substack{ \ON{k} = 0\\ \Delta_{ke} = 0 \\ (k \ne m,n) }} - \left. \rRRDM^{\DNO,\up\dw}_{pqpq} \right|_{\substack{ \ON{k} = 0\\ \Delta_{ke} = 0 \\ (k \ne m) }} - \left. \rRRDM^{\DNO,\up\dw}_{pqpq} \right|_{\substack{ \ON{k} = 0\\ \Delta_{ke} = 0 \\ (k \ne n) }}\right) 
		&, \text{if } m \ne n
	\end{cases}
\end{equation}
The $G^{\DNO,\dw\dw}_{mn}$ and $G^{\DNO,\dw\up}_{mn}$ terms are defined analogously.  
By zeroing the contributions of other electron pairs, the sum captures the contributions to the 2-RDM from the electron pairs that originate from the occupied NOs $m$ and $n$ only. 
When $m\ne n$, the intrapair contributions are removed.  
Also notice, the sum over 2-RDM elements is only over the Coulomb-like terms ($pqpq$) for $m\ne n$, and the time-inversion exchange and Coulomb-like terms for $m=n$.  
These definitions lead to the following expressions for the static correlation and high-spin correction contributions to the damping factors, for active-occupied spin-orbitals,  
\begin{equation}
	\beta_{m\bar{m}} =\beta_{\bar{m}n} = 1 + \sum_{pq} \sqrt{\Delta_{mp}\Delta_{mq}} - 2\sum_p \tau_{mp}
\end{equation}
\begin{equation}
	\beta_{m\bar{n}} = \beta_{\bar{m}n} = 1 + \zeta_{mn} - 4\kappa_{mn}
\end{equation}
\begin{equation}
	\beta_{mn} = \beta_{\bar{m}\bar{n}} = 1 + 4\kappa_{mn}
\end{equation}
In the case of virtual NOs, it is possible that occupancy is transferred from multiple occupied NOs.  
Therefore, the contributions of static correlation and the high-spin correction to the damping factor is combined through multiplication,
\begin{equation}
	\beta_{mf} = \prod_{\substack{ n \\ (\Delta_{nf} \ne 0)}} \beta_{mn}.
\end{equation}
The resulting contribution is a product of terms for the electron pairs that are transferred to that particular virtual. 
If both spin-orbitals are active virtuals then the product includes all factors for separate electron pairs that are transferred to those virtuals,
\begin{equation}
	\beta_{ef} = \prod_{\substack{ mn \\ (\Delta_{me} \ne 0) \\ (\Delta_{nf} \ne 0)}} \beta_{mn}.
\end{equation}
The vacancy-vacancy contributions are equivalent to the occupancy-occupancy terms,
\begin{equation}
	\beta^{pq} = \beta_{pq}.
\end{equation}
}

\josh{
The intermediate sums, $G^{\DNO,\up\up}_{mn}$ and $G^{\DNO,\up\dw}_{mn}$, are the contribution to the 2-RDM from a particular electron pair. Due to the pair-correction term of the $\DNO$ 2-RDM, the denominator of $\beta_{mn}$ [Eq.~\eqref{eq:betamn}] is unity,
\begin{equation}
	G^{(0),\up\up}_{mn} + \left( G^{\DNO,\up\up}_{\pair} \right)_{mn} = G^{(0),\up\dw}_{m\bar{n}} + \left( G^{\DNO,\up\dw}_{\pair} \right)_{m\bar{n}} = 1 .
\end{equation}
The expressions for $\beta_{mn}$ and $\beta_{m\bar{n}}$ can be rewritten accordingly,
\begin{align}
	\beta_{mn} & = 1 + \left( G^{\DNO,\up\up}_{\stat} \right)_{mn} + \left( G^{\DNO,\up\up}_{\HSC} \right)_{mn},
	&
	\beta_{mn} & = 1 + \left( G^{\DNO,\up\dw}_{\stat} \right)_{mn} + \left( G^{\DNO,\up\dw}_{\HSC} \right)_{mn}.
\end{align}
The expressions for the occupancy-vacancy damping effect, $\beta_m^n$ and $\beta_m^{\bar{n}}$, can also be written in the same form, however static correlation and high-spin correction have the opposite effect,
\begin{align}
	\beta_{mn} & = 1 - \left( G^{\DNO,\up\up}_{\stat} \right)_{mn} - \left( G^{\DNO,\up\up}_{\HSC} \right)_{mn}, 
	&
	\beta_{mn} & = 1 - \left( G^{\DNO,\up\dw}_{\stat} \right)_{mn} - \left( G^{\DNO,\up\dw}_{\HSC} \right)_{mn}.
\end{align}
This is due to the inverse nature of vacancy compared to occupancy (\textit{i.e.},~when the spin-up orbital is [locally] occupied the spin-down orbital is vacant, and vice versa). 
When applied to spin-orbital pairs involving virtuals, the effects from transferring different electron pairs to that virtual are combined via multiplication, 
\begin{align}
	\beta_m^f & = \prod_{\substack{ n \\ (\Delta_{nf} \ne 0)}} \beta_m^n,
	& 
	\beta_e^f & = \prod_{\substack{ mn \\ (\Delta_{me} \ne 0) \\ (\Delta_{nf} \ne 0)}} \beta_m^n.
\end{align}
The above definitions lead to the following expressions for the occupancy-vacancy contributions to the damping factors,
\begin{align}
	\beta_m^{\bar{m}} & = 1 - \sum_{pq} \sqrt{\Delta_{mp}\Delta_{mq}} + 2\sum_p \tau_{mp},
	& 
	\beta_m^{\bar{n}} & = 1 - \zeta_{mn} + 4\kappa_{mn},
	&
	\beta_m^n & = 1 - 4\kappa_{mn}.
\end{align}
}

\josh{
Finally, if all indices correspond to active spin-orbitals, then the damping factor is zero, \textit{i.e.},
\begin{align}
	\damp{mn}{ef} = \damp{\bar{m}\bar{n}}{\bar{e}\bar{f}} = \damp{m\bar{n}}{e\bar{f}} = \damp{\bar{m}n}{\bar{e}f} = \damp{\bar{m}n}{e\bar{f}} = \damp{m\bar{n}}{\bar{e}f} &	= 0, \quad \text{ if } \NO{m} \land \NO{n} \land \NO{e} \land \NO{f} \in \Act.
\end{align}
It is assumed that such interactions are already included in the $\DNO$ 2-RDM.
}

\bibliography{MP2_CCSD}

\end{document}